\documentclass[letterpaper]{article} % DO NOT CHANGE THIS
\usepackage{aaai25}  % DO NOT CHANGE THIS
\usepackage{times}  % DO NOT CHANGE THIS
\usepackage{helvet}  % DO NOT CHANGE THIS
\usepackage{courier}  % DO NOT CHANGE THIS
\usepackage[hyphens]{url}  % DO NOT CHANGE THIS
\usepackage{graphicx} % DO NOT CHANGE THIS
\usepackage{natbib}  % DO NOT CHANGE THIS AND DO NOT ADD ANY OPTIONS TO IT
\usepackage{caption} % DO NOT CHANGE THIS AND DO NOT ADD ANY OPTIONS TO IT
\usepackage{algorithm}
\usepackage{algorithmic}

\usepackage{subcaption}

\usepackage{newfloat}
\usepackage{listings}
\DeclareCaptionStyle{ruled}{labelfont=normalfont,labelsep=colon,strut=off} % DO NOT CHANGE THIS
\floatstyle{ruled}
\newfloat{listing}{tb}{lst}{}
\floatname{listing}{Listing}
\title{Accountability Capture: How Record-Keeping to Support AI Transparency and Accountability (Re)shapes Algorithmic Oversight}
\author{
    Shreya Chappidi\textsuperscript{\rm 1,2}, 
    Jennifer Cobbe\textsuperscript{\rm 1},
    Chris Norval\textsuperscript{\rm 3},
    Anjali Mazumder\textsuperscript{\rm 4},
    Jatinder Singh\textsuperscript{\rm 1,5}
}
\affiliations{
    \textsuperscript{\rm 1}University of Cambridge \\
    \textsuperscript{\rm 2}National Institutes of Health \\
    \textsuperscript{\rm 3}University of Aberdeen \\
    \textsuperscript{\rm 4}The Alan Turing Institute \\
    \textsuperscript{\rm 5}Research Center Trustworthy Data Science \& Security, University Alliance Ruhr \\%
}

\usepackage{bibentry}
\begin{document}

\maketitle


\maketitle

\begin{abstract}

Accountability regimes typically encourage record-keeping to enable the transparency that supports  oversight, investigation, contestation, and redress. However, implementing such \textit{record-keeping can introduce considerations, risks, and consequences, which so far remain under-explored.} This paper examines how record-keeping practices bring algorithmic systems within accountability regimes, providing a basis to observe and understand their effects. For this, we introduce, describe, and elaborate \textit{`accountability capture'} -- the re-configuration of socio-technical processes and the associated downstream effects relating to record-keeping for algorithmic accountability. Surveying 100 practitioners, we evidence and characterise record-keeping issues in practice, identifying their alignment with accountability capture. We further document widespread record-keeping practices, tensions between internal and external accountability requirements, and evidence of employee resistance to practices imposed through accountability capture. We discuss these and other effects for surveillance, privacy, and data protection, highlighting considerations for algorithmic accountability communities. In all, we show that implementing record-keeping to support transparency in algorithmic accountability regimes can itself bring wider implications -- an issue requiring 
greater attention from practitioners, researchers, and policymakers alike.

\end{abstract}

\section{Introduction}
\label{sec:introduction}

There is growing attention and critique towards how algorithmic (and `AI') systems are commissioned, designed, deployed, and used.
Laws such as the European Union's (EU) General Data Protection Regulation (GDPR) ~\citep{gdpr2016}, AI Act~\citep{EUAIAct}, and Digital Services Act~\citep{dsa2020} provide for \textit{accountability regimes} to interrogate and regulate the use of digital technologies. Regulators and policy-makers have called for better oversight of how systems are designed and used~\citep{EuDenmark, canadaAlgAssess, commons2018, koene2019, ico2019, denmarkLaw, ico2020}. {Accordingly,} academic researchers have proposed frameworks of  explainability~\citep{arrieta2020}, reviewability~\citep{cobbe2021}, contestability~\citep{kaminski2021}, traceability~\citep{kroll2021}, and related approaches. Civil society groups have similarly proposed auditability, transparency, and accountability frameworks~\citep{adalovelace2020, demos2020, ainow}.

Record-keeping is key to accountability regimes, producing the information necessary for supporting meaningful transparency. \textit{Record-keeping practices} involve storing and managing information relating either to the technical components of algorithmic systems (e.g., logs describing system operation) or to broader human and organisational processes around them (e.g., {their} commissioning, design, deployment and use). 
Such information, in turn, can assist accountability, being provided to support understanding, scrutiny, investigation, oversight, contestation, and redress by a range of stakeholders~\cite{wieringa2020,cobbe2021}.

Importantly, technical and organisational record-keeping processes relating to algorithmic systems may require {institutional} changes (reconfiguration) to ensure that relevant information is generated or recorded. In turn, {these record-keeping processes} might change how an organisation's employees engage with the accountability regime as they re-orient to navigate and (potentially) evade it. Records will also often collect information about people using or subject to systems, implicating surveillance, privacy, and other legal issues. Further, technical mechanisms supporting record-keeping must accord with the needs of varying audiences, including regulatory bodies and organisations. 
{As such, the design of record-keeping mechanisms is both shaped by, and influences, emerging accountability regimes, and can produce downstream effects on actors and organisations.} 

\subsection{Motivation and Contributions}
There are growing legal and regulatory\footnote{
While we mainly refer to EU law, the record-keeping issues and their effects apply broadly across jurisdictions.} 
pressures for enhanced record-keeping, echoed across responsible AI standards and best-practice documents.
However, the effects and implications of accountability-related record-keeping are so far under-considered. 
Current record-keeping discussions primarily relate to transparency---focusing on the information that might support accountability or audit, and how to best obtain, provide, and understand 
that information~\cite{norval2022, cobbe2021, kroll2021, Wieringa_2020, Casper_2024, Simkute_2021, Birhane_2024}---rather than the processes effecting and governing record-keeping practices themselves. Here, we \textit{highlight and analyse the real-world impact of record-keeping practices}, and emphasise the associated risks for those devising accountability regimes or implementing record-keeping practices.

Specifically, this paper \textbf{explores the implications, challenges, and risks around accountability-oriented record-keeping}. {To provide a theoretical foundation,} we  draw and build from Agre's concept of `\textit{capture}'~\citep{agre1993}, which describes how human activity is reconfigured to become amenable for tracking by and with computers. Extending this concept of `capture' to the process of designing and implementing record-keeping practices for accountability purposes, our paper contributes the following:
\begin{enumerate}
    \item 
    
{Establishes the concept of \textbf{`accountability capture':} how reconfiguring algorithmic systems for accountability record-keeping requirements can have potentially wider and unintended effects.}
\item {Provides evidence from a survey of 100 professionals, indicating record-keeping practices and the emerging dynamics of accountability capture.}
\item {Applies the lens of accountability capture to key considerations around surveillance and privacy, to highlight emerging risks for those shaping, deploying, or regulating record-keeping systems.}
\end{enumerate}

\section{Information for Supporting Accountability}
\label{sec:accountoverview}

{In this section, we describe accountability and discuss how record-keeping practices help support it.}
\subsection{Defining Accountability}

{Definitions and conceptions of accountability vary across institutions and domains as governance structures are rapidly adapting to widespread algorithmic deployment \cite{Cobbe_Singh_2024, Williams_moving_2022}.} When discussing accountability around algorithmic systems, we, {like others~\citep{wieringa2020, cobbe2021, Binns_2018}}, draw from \textit{Bovens' model of accountability}~\citep{bovens2006}.
In Bovens' model, accountability involves an \textit{actor} from whom accounts are owed, a \textit{forum} to whom those accounts are owed, and a \textbf{relationship} of accountability between them. Here, actors provide information (\textit{accounts}) to forums, to support \textit{(i)} review, investigation, and oversight, and \textit{(ii)} effective and appropriate responses by the forum. 

Legal, regulatory, institutional, or organisational accountability regimes 
establish accountability relationships by identifying actors from whom accounts are owed; activities about which accounts are owed; and forums to whom they are owed. {Regimes can reflect more or less formalised prescriptive frameworks from either internal (e.g., organisational codes of conduct) or external (e.g., regulatory reporting requirements) forums requiring account-giving.}
These accounts can enable audit and scrutiny, which is increasingly discussed regarding algorithmic systems~\citep{chock2022,raji2022}. 
{Thus, under Bovens' model of accountability, \textit{any} form of record-keeping can be potentially understood as supporting some form of accountability where its records enable or support the provision of accounts to a forum.}

{Accountability regimes may be imposed either \textit{externally} (such as through law, regulation, or policy) or \textit{internally} (such as by organisations themselves seeking to better understand their own systems). Generally, algorithmic accountability regimes seek information about \textit{(i)} actor organisations (e.g., their organisational processes and technical logs and details); and \textit{(ii)} the activities that actors organisations enable (e.g., actions on their application, service or platform). 
{As algorithmic accountability requirements become more formalised, internal and external accountability demands are likely to overlap; e.g., when in-house legal teams align company practices with external audits or regulatory guidance.}
{As such, current record-keeping practices are relevant for shaping proposals and understanding the impact of evolving accountability regimes.}

\subsection{The Role of Records} \label{sec:roleofrecords}

Accountability---providing meaningful accounts to support understanding and action by forums---{often necessitates the keeping of \textbf{records}}~\citep{cobbe2021} describing the design, deployment, or operation of organisational and socio-technical systems. This requires that actor organisations employ \textit{record-keeping practices}: 
technical or organisational measures to collect appropriate records, providing information that enables the meaningful transparency needed to support accountability regimes. However, as we argue, record-keeping \textit{itself} can bring positive or negative implications that may require action to support or mitigate.

\subsubsection{Supporting Legal and Regulatory Accountability Relationships. ~} 

Accountability-related record-keeping requirements are often imposed by law to assist in understanding system design and behaviour.
The EU (GDPR){, for example,} requires those processing personal data keep records of that processing and potentially make them available to regulators and others~\citep[Art. 30, recital 82]{gdpr2016}. 

Given the recent increased availability of algorithmic tools and services, the EU's Digital Service Act contains powers for judicial and administrative authorities to obtain records from organisations relating to users of their services~\citep[Art. 10]{dsa2020}. The EU's Artificial Intelligence Act requires those responsible for `high risk' AI systems to maintain records relating to their system's development, capabilities, and limitations~\citep[Arts. 10-11, 13, 18]{EUAIAct}, including around processes for checking models and datasets for biases and errors, and to implement automated logging mechanisms for deployed systems (Arts. 12, 20). Moreover, records may also assist {with human oversight activities, such as those specified by the EU AI Act~\citep{sterz_quest_2024}, as well as} regulatory\slash compliance management and determining liability when harm arises.

\subsubsection{Supporting Other Types of Accountability Relationships. ~} 
Records about algorithmic systems thus potentially support many accountability relationships and regimes, including regulators and oversight bodies through compliance and enforcement actions or civil society through scrutiny and advocacy. Records may help individuals using or affected by systems to be better informed and to challenge an actor and its systems. 
Record-keeping can help actors developing algorithmic systems with assessing their systems, workflows, and processes in line {with} organisational requirements and standards  (`bureaucratic accountability'~\cite{williams2021}). 
When incidents occur, records can assist with repairing, debugging, and mitigations to prevent recurrence, reflecting internal accountability regimes established to support product design and operation.

\subsection{Implementing Record-Keeping Practices}

Instituting record keeping practices for an accountability regime involves considering \textit{(i)} what information is required, and \textit{(ii)} how the algorithmic systems need to be developed and configured to best support record-keeping.

\subsubsection{Two Kinds of Record-Keeping.~ }
Broadly speaking, record-keeping practices take two forms: records \textit{from} and \textit{about} an algorithmic system. \textbf{Records from the system} describe how the system and its workflows actually functioned. These records may include how systems were used, details of actions taken through interfaces, % taken, data accessed, and services used, as well as the
and results of any processing or models employed. Meanwhile,  \textbf{records about the system} concern how an algorithmic system was commissioned, built, deployed, and monitored. Typically, these details include documents, notes, training and testing datasets, validation criteria, risk assessments, %development frameworks, 
model specifications, staff training manuals 
%and guidance
, etc.~\cite{cobbe2021}. In this way, some records concern what \textit{should} occur, and others what \textit{does} occur; and therefore can include information about both `design-time' and `run-time' aspects of systems.

\subsubsection{Information Requirements.~ } 
Accountability is inherently \textit{contextual} -- the information and corresponding records needed to support accountability depend on the actors, forums, and accountability regimes involved~\citep{bovens2006, stohl2016,Ananny_Crawford_2018, Cobbe_Singh_2024}. 
Accountability relationships between actors and regulators (`administrative accountability'), for instance, differ from those between actors and professional peers (`professional accountability')~\citep{williams2021}. There may also be contextual overlaps or differences between internal standard organisational practices and requirements implemented by external accountability regimes. Contextually appropriate information is: \textit{relevant} to the accountability relationships; \textit{accurate} in that it is correct, complete, and representative; \textit{proportionate} to the level of transparency required; and \textit{comprehensible} by forums~\citep{cobbe2021}.

\subsubsection{Technical Configuration.}
Record-keeping systems will need to be designed, instrumented, adjusted, or extended to enable the technical recording of relevant information and support the broader organisational record-keeping practices. Record-keeping will often entail both \textit{passive} measures (automatic mechanisms recording details of particular system events, inputs, user actions, etc.) and \textit{active} measures (procedures requiring human input to record information not captured automatically, e.g., reasons behind user actions such as GitHub commit messages).

% \textit{active} measures (less automated procedures  %recording information that may be invisible to automated record-keeping mechanisms). Active measures {(e.g., github commit messages)} 
% that may capture information about reasons behind user actions (e.g., github commit messages)). 

% such as having them describe why particular data was accessed or why an algorithmic decision was overruled. 

Various recent approaches for algorithmic record-keeping propose recording information about particular aspects of the algorithmic process (though holistic approaches are emerging including `reviewability'~\citep{cobbe2021} or `traceability'~\citep{kroll2021}). For instance, `datasheets for datasets'~\citep{datasheets2018} and other works \cite{Hutchinson_2021, Werder_Ramesh_Zhang_2022} propose recording information about the quality of datasets used in model training, while `model cards'~\citep{mitchell2019model} and `factsheets'~\cite{sokol2020factsheets} propose recording key model details about. Emerging organisational measures include standards, questionnaires for identifying sources of unintended algorithmic bias~\citep{lee2021}, and auditing frameworks ~\citep{icoauditing}. {While we are not concerned here with such proposals directly, these illustrate trends in record-keeping practices suggested in literature.}

\section{Effects of Record-Keeping: 
Accountability Capture}
\label{sec:capture}

{Exploring the impacts of record-keeping for accountability requires inquiry into technical and organisational record-keeping processes, resulting human behaviors, and associated requirements imposed across accountability regimes.}
To analyze record-keeping and associated effects, we apply the lens of \textit{capture}~\citep{agre1993} which 
{recognises that computerised tracking of human activities requires those activities to be (re-)configured through a complex socio-technical process that reflects underlying logics of computerisation}. Drawing from this, we argue that record-keeping can produce \textit{accountability capture}, bringing downstream transparency
implications for actor organisations and others.

\subsection{Agre's Theory of `Capture'}

Agre derived his theory of \textit{capture} in the 1990s in the context of increasing computerisation. Capture is described as ``the deliberate reorganization of industrial work activities to allow computers to track them in real time''~\citep{agre1993}. Capture involves human and organisational processes (e.g., managing supply chains) being reconfigured for computerised tracking, permitting control for purposes such as understanding workflows and optimising systems. Indeed, the re-configuration of organisational activities around logics of computation is a key feature of capture. Computers do not simply record neutral information, instead recording \textit{representations} of activities using data shaped by human choices about what can and should be represented computationally.

Agre identifies five phases of the capture process, {drawn from observations of real-world computerised tracking}: 

\noindent \textbf{1. Analysis}:
{Studying an activity and conceptualising its fundamental units for representation by computers.}
    
\noindent \textbf{2. Articulation}: 
Articulating \textit{grammars of action} describing how the fundamental units identified in the `analysis' phase can be ``strung together'' to form ``stretches of activity'' that facilitate computerised tracking.
    \\
\noindent \textbf{3. Imposition}: 
Re-configuring activities in line with the grammars of action by inducing people engaged in the articulated activity to act accordingly. 
    Imposition is typically social (procedures involving some relations of authority) and technical (involving machinery or physical barriers).
    \\
\noindent  \textbf{4. Instrumentation}: 
    Providing organisational and technical means to collect information about the ongoing re-configured activity for use in computational processes for control. Participants in the activity ``begin, of necessity, to orient their activities towards the capture machinery and its institutional consequences.''
    \\
\noindent  \textbf{5. Elaboration}: 
    Storing, inspecting, auditing, and analysing the captured activity's records, merging them with other records, using them for optimisation, further calibrating the grammars of action and captured activity, and so on. 

Through these phases, human activities can be brought within the reach of automated tracking by computers.

\subsection{Accountability Capture}

While Agre focused on `capture' as it relates to organisational gains (and did not discuss accountability or algorithms directly), we argue that {designing and implementing record-keeping practices to meet} accountability regimes {reflects and builds on capture's phases.} 
{During this process, algorithmic and organisational systems are created or (re)-configured to embed accountability requirements that produce or retain information suitable for account-giving.} Reflecting on Agre's description of capture, we describe and argue that \textbf{accountability capture}---or capture produced through record-keeping intended to meet accountability requirements---operates through analysis, articulation, imposition, instrumentation, and elaboration to make human, organisational, and socio-technical processes amenable to the control of an accountability regime.
{While Agre's traditional conception of capture refers to reconfiguration of activities brought by computerised tracking of human behavior, accountability capture emphasizes the `capture within captured' that occurs when record-keeping processes shape and re-orient socio-technical processes and data captured on surrounding algorithms that {may} already {be} gathering and processing data on human behaviors themselves.}

\subsection{Case Study: Worker Monitoring Algorithms}

Organisations are increasingly implementing algorithms to monitor, evaluate, and compare employee performance, using metrics such as compliance with hybrid work policies requiring in-person employee attendance \cite{Ajunwa_2019, Aloisi_Gramano_2019}. Many jurisdictions, including the EU's Charter of Fundamental Rights (Title IV, Art. 30) protect against employee termination without cause. Moreover, some data protection legislation explicitly contains employment-related provisions, such as GDPR [Art. 88] enabling rules to protect ``rights and freedoms in respect of the processing of employees’ personal data in the employment context'' \cite{Aloisi_Gramano_2019}. External oversight from data protection and employee protection bureaus may influence internal activities from legal and HR departments, necessitating record keeping on ``evidence-based'' employee monitoring practices \cite{Aloisi_Gramano_2019}.
{This section uses a case study---grounded in reporting on hybrid work and employee monitoring \cite{Owl_Labs_2023}---to illustrate ‘accountability capture’ in a domain subject to regimes with varying levels of formalisation (e.g., internal legal teams and external employment and data protection actors).}}

\subsubsection{Components of Algorithmic Record-Keeping.} First, the process of deciding which  aspects to document about the employee monitoring algorithm involves \textit{analysis} of the purpose involved in establishing the algorithm. This might include review of legal hiring and termination requirements, formalisation of employee job descriptions, key performance indicators (KPI), and organisational reflections on the ideal hybrid employee's performance. Then, the \textit{articulation} phase involves the stringing together of these components identified as relevant to establishing the employee monitoring algorithm into `grammars of action' that can be tracked via computers. For example, \textit{articulation} might involve scrutiny over the legal, organisational, and technical arms within scope during the creation, justification, and operation of the algorithm. The process may also include details on how to measure the relevant units identified during \textit{analysis}, including how precisely or frequently to track job expectations (e.g., on-time deliverables and other KPIs) or hybrid work policy compliance (e.g., `badging' frequency of tapping their card to physically enter/exit the workplace). 

\subsubsection{Implementing Algorithmic Record-Keeping.} The record-keeping process could inform other actors as to the nature of the algorithmic monitoring during the latter accountability capture phases. During \textit{imposition} of the algorithmic record-keeping, contractual language may be inserted into employment agreements that authorises employee monitoring via certain metrics (e.g., laptop and network access). During \textit{instrumentation}, IT or HR specialists may be asked to create documentation on the employee monitoring algorithm and {associated features implemented to enable monitoring}. Lastly, \textit{elaboration} may entail the analysis and updating of records related to thresholds and features captured within the algorithmic monitoring system. 

\subsubsection{The Effects of Accountability Capture.~ } The end result of accountability capture may involve behavior change from affected employees who gain insight into monitoring algorithms during the implementation of record-keeping. {Consider, for example, the case where there is} employee dissatisfaction with return to office mandates and {performance} monitoring \cite{Parkinson_Schiavon_Kim_Betti_2023}. The process of keeping records on monitoring algorithms may facilitate employee investigation into evolving algorithmic metrics or thresholds employed to issue warnings (e.g., requiring x of badge-ins over y days). {As a result of record-keeping \textit{imposition} and \textit{instrumentation} during accountability capture,} employees may re-orient to evade these sanctions owing to records imposed by accountability regimes, such as through `coffee badging' or briefly badging in to the office and leaving \cite{Owl_Labs_2023}. Meanwhile, the capture apparatus may be updated through the accountability regime, such as Amazon re-defining an office visit duration to minimum 2 hour periods \cite{kim_2024}, requiring new social imposition and technical instrumentation (e.g.,  adding badge-\textit{out} tracking).

\subsubsection{Summary.} The framework of accountability capture specifically considers amongst other things
the creation of records from and about systems that are
tracking human behavior (e.g., attendance), allowing for the
analysis of scenarios of ‘capture within the captured’
(e.g., investigation into established measures and metrics). Thus, accountability capture operates to reconfigure practices around algorithmic systems (e.g., updating the record-keeping frequency or documentation format of a monitoring system) that are themselves reconfiguring practices (e.g., employee behavior).

This case study highlights the mixed implications of `accountability capture,' where {competing internal and external factors, such as} legal record-keeping requirements through employment and data protection regulations, organisational motives, and employee rights and incentives may influence the process of algorithmic record keeping, including related frequency, documentation, and implementation methods. As such, just as Agre discussed how `computerisation' generally entailed capture, {we argue that} accountability regimes may similarly provoke, necessitate, shape, or intensify the capture of human and organisational processes \textit{and} the activities of technical components in various different ways {to produce} \textbf{accountability capture}.

\section{Practitioner Survey on Record-Keeping}
\label{sec:survey}

To understand how accountability capture arises {(or could arise)} in practice, we surveyed 100 professionals managing, building, or using algorithmic systems. Through this, we \textit{(i)} explore current record-keeping practices of organisations using algorithmic systems, and \textit{(ii)} outline real-world effects of accountability capture faced by participants {from internal and/or external accountability regimes.} 
{The aim of the survey was to provide empirical grounding for the concept of accountability capture by examining how record-keeping approaches, their drivers, and their consequences play out in practice, and to identify areas for further attention as accountability regimes become more formalised.}

\subsection{Survey Method and Details}

Our survey asked questions aimed to investigate (i) broad record-keeping practices, (ii) the specific practices implemented (if any) to meet internal and/or external accountability regimes (e.g., legal/regulatory drivers), {and (iii) the associated effects of implementing such record-keeping.}\footnote{Full survey data and code used to generate figures are available at \url{github.com/shreyarch/accountabilitycapture}.}

Through the Prolific platform~\cite{prolific2023}, we initially recruited 400 working professionals with experience in software development. {Participants completed a brief pre-screening survey, and those with self-reported experience working with or managing algorithmic systems were invited to complete the full survey  until we obtained our desired sample size of 100, broadly in line with other studies engaging online qualitative feedback~\cite{caine2016}.} 

Participants were recruited from 21 countries across 4 continents (Fig. S1a), representing a wide range of sectors including \textit{commerce}, \textit{professional services}, and \textit{finance} (Fig. S1b). Participants held a diverse array of roles, including developers, consultants, and data analysts/scientists (Fig. S1c), with most holding 0-5 years of experience (Fig. S1d). 
We obtained prior approval from our institution's ethics review board, and all respondents were compensated in line with the UK's `living wage' {\cite{HM_Revenue_Customs}}.

Percentages reported in the following section reflect answers to multiple choice questions (generally with multi-select enabled), while individual numbers reported (e.g., n=8) refer to common responses grouped during thematic analysis of optional free text responses.

\subsection{Accountability Capture in Practice}

We now describe the extent to which accountability capture {occurs or could be observed during the} creation and imposition of algorithmic record-keeping practices using real-world evidence captured in our survey. 
{Ongoing organisational practices form the baseline for understanding the realised and potential impacts of accountability capture, as future accountability-oriented record-keeping requirements will likely be implemented in these contexts. As such, we examined and categorised the diverse purposes and methods of record-keeping, distinguishing where internal and external accountability regimes played a role.} Bolded texts represent summarised key findings relating to the phases of accountability capture identified within the survey.

\subsubsection{(1) Analysis.~ } 
\label{sec:survey-analysis}
In the frame of accountability capture, \textit{analysis} might involve reviewing the design and functionality of an algorithmic system or its associated accountability mechanism, followed by a review of which algorithmic aspects can be documented or recorded. Key observations may include actors, resources, and stages involved with creation, tooling, or application of an algorithmic system throughout the development lifecycle \cite{De_Silva_Alahakoon_2022}.

\textbf{Algorithmic systems are being widely deployed across sectors and organisational processes, {involving forums with diverse information requirements.}}
Surveyed organisations used algorithmic systems for diverse purposes, including management, hiring/recruiting, purchasing/finance, and other logistics/supply chain mechanisms. Participants often reported multiple parties responsible for operating these algorithmic systems, including in-house actors (91\% of responses), other businesses (39\%), consumers (19\%), and external government, public sector, or charity organisations (14\%). Use of these algorithmic systems often impacted multiple parties, including employees (88\% of responses), people from other organisations (41\%), and other individual system users such as customers (48\%), indicating widespread potential for accountability capture effects.

\textbf{The algorithmic system design and development processes vary widely between organisations.} \textit{Analysis} likely requires a detailed understanding of the conceptualisation, design, implementation, and operation of the algorithmic system in question. {Analysis is central to Bovens' definition of accountability, as an understanding of the system is required to better enable appropriate account-giving to a forum, who may have different information requirements based on their contextual needs.}
Surveyed algorithmic systems often involved multiple parties in their design, including in-house individuals (65\% of responses), contractors or consultants (41\%), organisations hired to build custom systems (32\%), commercial systems for multi-purpose use (30\%), or with open-sourcing (17\%). These results indicate a wide array of actors called upon during the design and development of algorithmic systems, who may hold varying motives, viewpoints, and levels of power towards actually shaping processes during the \textit{articulation} stage.

\subsubsection{(2) Articulation.~ } 
\label{sec:survey-articulation}

Agre's second stage of capture involves \textit{articulation}, or development of grammars of action that combine units identified in the prior \textit{analysis} phase and support computerised tracking. In the context of accountability capture, this may include organisation-level decisions of how different aspects of the algorithmic system interact and how data on such interactions should be captured.  

\textbf{Record-keeping is widespread.} Results indicated that organisations have largely engaged in the \textit{articulation} process as 87\% of participant organisations automatically record information about their system. These automatic records were broadly reported to include customer information, usage logs\slash tracking, evaluative metrics, and operations\slash financial details. 
Individuals within the organisation (88\%) were largely involved in defining and articulating organisational record-keeping specifications. These responsible parties included management (n=48), specialized departments including audit, fraud, and legal (21), and project manager/software engineer teams (18).  Only 15\% of participants agreed that their organisation was not recording information that it should or would prefer to record, indicating relatively high satisfaction with the overall \textit{articulation} process. {These results evidence the ongoing, widespread process of record-keeping across diverse organisations, indicating the need for stakeholders calling for accountability to study and be aware of existing record-keeping practices. }

\textbf{Record-keeping practices are guided by both internal and external organisations.} At the same time, the \textit{articulation} process is shaped by input from multiple stakeholders. Participants reported engaging internal guidance (e.g., ranking the importance of data to collect) and 58\% previously sought external guidance from parties such as regulators, industry bodies, and consultants about their record-keeping processes. 16\% also explicitly indicated that external actors were \textit{primarily} responsible for defining their organisational record-keeping processes, demonstrating the already influential role of external entities {in establishing broad explicit and/or sector-specific accountability requirements.}

\textbf{Designing computerised record-keeping systems is an imperfect process, with many records persisting in manual formats.} {As Agre argues,} grammars of action breaking down activities into component units may oversimplify or distort captured activities ~\citep{agre1993,poirier2022accountableData}. Consequently, the activity’s `invisible' aspects may be unrepresented, while the grammars of action may elide aspects of how activities interweave, or mistake prescribed procedures for an accurate account of the activity in practice~\citep{agre1993, poirier2022accountableData}. Our results also indicated a varied process of articulation, with 47\% of participant organisations continuing to engage in manual record-keeping processes (as opposed to fully automated tracking). Manually recorded information included system design \& updates (n=15), personal data (11), organisational information (9), debugging / error logging (4), customer feedback (3), and performance statistics (2). Reasons for manually recording information included needs for system documentation (n=15), general operations requirements (15), performance evaluation desires (8), organisation requirements (3), regulation (1), desire for system backup (1), and lack of automated system abilities (1). These results may indicate challenges in the analysis or articulation phase for more complex components of record-keeping, such as the establishment of sufficient documentation, that supports continued manual records. {The manual or automatic units identified during articulation affect the latter stages of accountability capture, 
and thus reflect the relevance of organisational structures and {technical design choices for algorithmic accountability in practice.}}

\subsubsection{(3) Imposition.~ } \label{sec:survey-imposition}

Agre's third stage of \textit{imposition} during capture involves the re-configuration of existing activity to align with grammars of action identified during \textit{articulation}. In the context of accountability capture, this might involve organisational and/or individualised actions to facilitate, implement, and communicate record-keeping practices. 

\textbf{Organisations primarily rely on internal or in-house actors to establish, articulate, and maintain record-keeping processes.} 
82\% of organisations had at least one in-house responsible party involved in ensuring appropriate compliance with record-keeping processes, with these generally including senior management and C-suite executives, as well as dedicated audit/legal, security, and/or information technology departments. 
Reported strategies to articulate, maintain, and verify these record-keeping processes included training, documentation, system design, organisational guidelines, workshops, testing, communication, reconciliation reports, periodic reviews, and quality indicators. These diverse approaches to establishing record-keeping requirements reflect the potential influence of organisational norms, practices, and management towards shaping the algorithmic capture apparatus. Additionally, they reflect the potential risks for the algorithmic capture apparatus to be defined and managed by a singular or particular perspective.

\textbf{Imposition often involves multiple stages of activity re-configuration {to support accountability regime requirements}.} 
When discussing barriers to automatically recording information, {participants explicitly discussed legal compliance related to privacy and consent (n=10), planning and execution difficulties (4), fairness / ethics considerations (3), maintaining only relevant information (2), performance errors (2), and labor displacement (1). 
Strategies suggested to overcome these challenges included communication explaining the accountability-related purposes for record-keeping data (n=7), dedicated planning \& specification strategies (5), onboarding and testing (4), stakeholder engagement research (4), establishment of an in-house ethics committee, and even lobbying of institutions. Moreover, at least 19\% of surveyed organisations implemented record-keeping mechanisms into already deployed algorithmic systems, adding additional records of performance metrics (n=8), system documentation (5), organisational reports (3), financial details (4), and customer analytics (3). These results indicate the complex, contextual, and fluid process of \textit{imposition} during capture, often requiring multiple forms of re-configuration to align record-keeping practices with the goals of the accountability regime.

\textbf{Competing notions of accountability often shape the record-keeping implementation process.} Sometimes the process of imposition can also put various stakeholders at odds, with 16\% of surveyed participants indicating that their organisation needs to record certain information (e.g., for legal or contractual purposes) that it would prefer not to be recording (Fig. \ref{fig:records-likert}). {People overseeing regimes are often not the people designing, implementing, or complying with record-keeping. As a result, there may be discrepancies between what those designing the regime expect to happen, what those designing record-keeping interpret the regime to mean should happen, and what occurs in practice. This may be relevant where regulatory external accountability regimes are developed in the abstract and require translation to specific record-keeping practices by actor organisations.} 

\begin{figure}[!t]
    \begin{centering}
\includegraphics[width=.85\columnwidth]{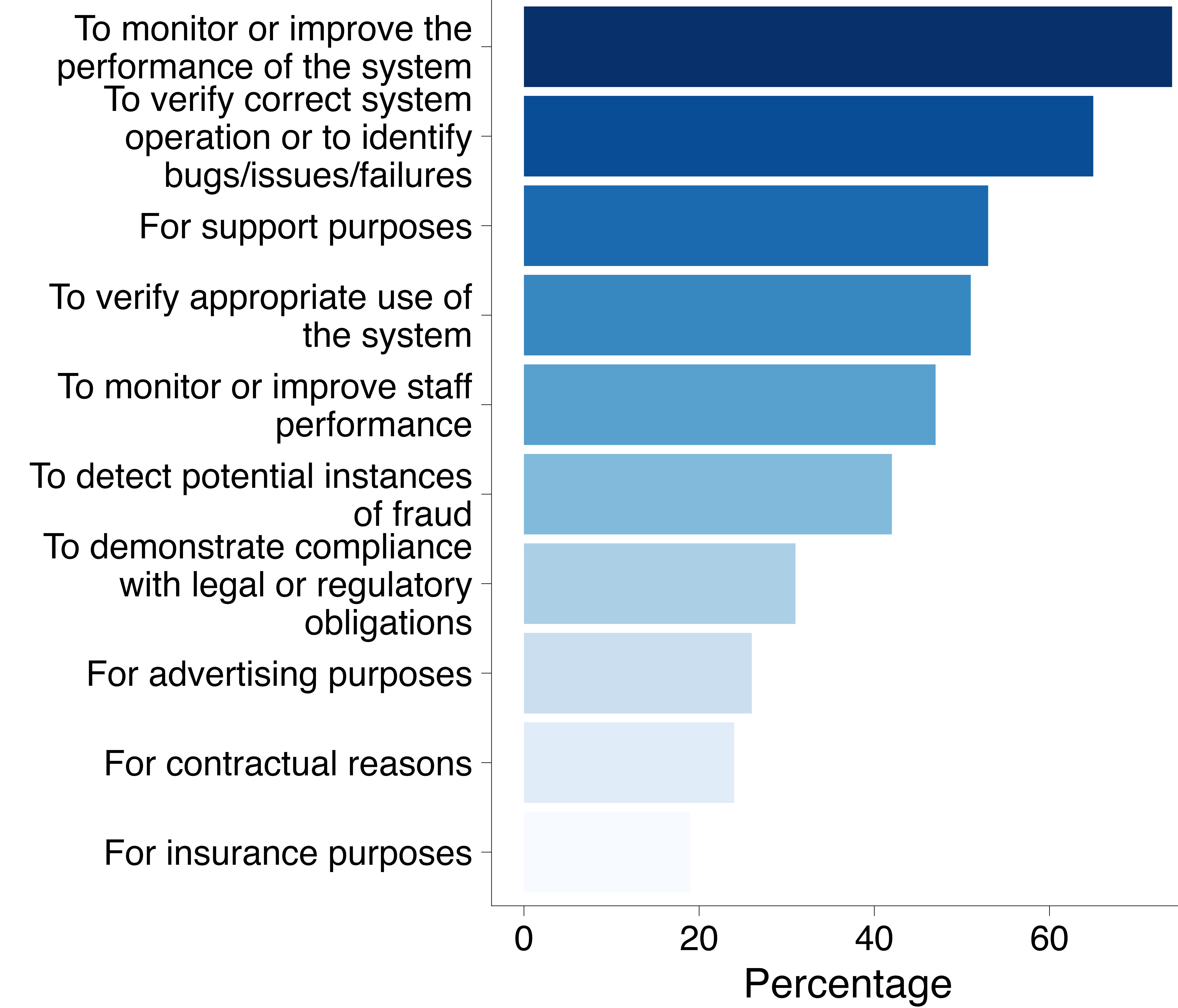}
\caption{Drivers for organisational record-keeping.}
    \label{fig:records-reasons}
% \end{centering}
% \end{figure}

% \begin{figure}
%     \begin{centering}
\includegraphics[width=.95\columnwidth]{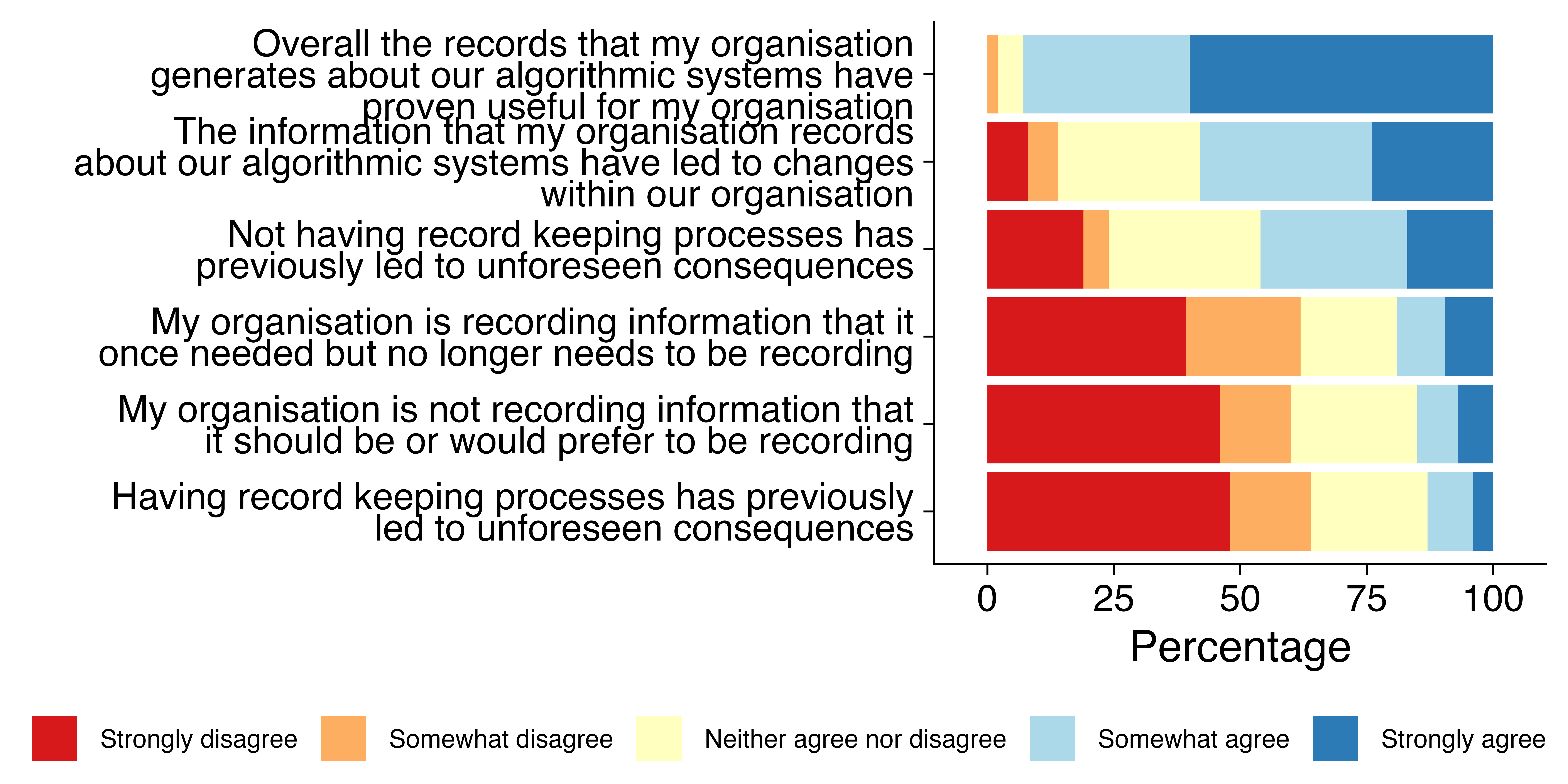}
    \caption{Organisational effects of imposed record-keeping.}
    \label{fig:records-likert}
% \label{fig:reasons}
% \vspace{-0.55cm} 
    \end{centering}
\end{figure}

\subsubsection{(4) Instrumentation.~ } \label{sec:survey-instrumentation}

Instrumentation refers to the organisational and technical processes of imposing the re-configured activities for computerised tracking, and involves participants orienting their activity towards ``the capture machinery and its institutional consequences.'' Through accountability capture, instrumentation might include deployment of automated record-keeping mechanisms and associated actor behavior in response to these novel mechanisms.

\textbf{{Instrumentation is influenced by both the actors and motives involved during the creation and implementation of algorithmic systems.}} Records kept about algorithmic systems as they were built included operation / use information (n=13), design / conceptualisation details (9), system upgrade history (8), performance metrics (8), error logging (6), system requirements (5), customer analytics (2), safety and efficiency guardrails (3), and data sources (1). The variety of records kept indicate diverse conceptions of an algorithmic system, its purpose, its constituent parts, and its relevance to an organisation. Yet, technical barriers to automated record-keeping remain. 13\% of participants who didn't automatically record information about their algorithmic systems cited personnel access restrictions and manual data processing methods, while those automatically recording data still cited issues of legal compliance (n=10), planning and execution challenges (4), fairness and ethics concerns (2), and concerns over labor displacement (1).

\textbf{Record-keeping produces changes in actor organisations and employee behaviours.} We found that 58\% of respondents agreed that record-keeping practices produced changes in their organisation, with the vast majority of these (50 out of 58) indicating that changes were positive (Fig. \ref{fig:records-likert}). Changes included increased compliance (n=7), employee wellbeing (5), efficiency (4), and fairness (1). 22\% of all respondents indicated that employees changed their behaviour following the introduction of record-keeping. In many cases, this was viewed as a positive change to support or engage with the accountability regime. One participant noted that employees ``became more serious about the need for documenting what is relevant,'' while another said that employees became ``more careful when it came to recording their order and customer details'' and were ``thorough when carrying out administrative tasks.'' These responses suggest that the phases of capture can make a positive difference by orienting actor internal activities towards accountability regimes.

\textbf{Record-keeping produces changes in behavior outside of the organisation.} Record-keeping may also elicit changes external to the organisation implementing such process, with 20\% of respondents indicating that internal algorithmic record-keeping practices generated changes outside of their own organisation, including the practices of clients, competitors, other cooperating businesses, and overall industry insights. Some of these changes included improved organisational optics, increased reporting from users, and increased financial investment. This reflects the wider scope and impact of accountability capture, given that the record-keeping process can influence industry trends and behavior of users subject to algorithms.

\textbf{Instrumentation of record-keeping practices may elicit \textit{resistance} from those subject to accountability regimes.} Encouraging changes in actors’ algorithmic processes is often part of an accountability regime's goal. Yet, resulting changes may not always be desirable to the party seeking accountability. Accountability capture can thus provoke forms of \textit{resistance}~\citep{foucault1990} to the capture process, which may undermine the accountability regime. 

Almost a quarter (n=23) of respondents said that employees have attempted to push back against or evade the record-keeping apparatus. In some cases, employees ``became less active with their work,'' and ``became depressed and less focused.'' Another respondent indicated that ``the biggest challenge is to push everyone to comply with the record-keeping practices.'' The extent to which employees seek to undermine record-keeping may also depend on factors such as their understanding of the technology: ``some employees are not skilled when it comes to using technology systems so they tend to evade some of the record keeping processes.''

This is particularly relevant as actors may appear compliant whilst actually undermining the regime’s efficacy. Participants in captured activities may, for instance, advantageously adjust the timing and contents of records; interpret and categorise events sympathetically; bias judgement calls; minimise interactions with activities that are recorded; include, delay or fail to enter certain information; or even falsify records. {Some respondents noticed ``resistance to change'' when record-keeping processes were brought in, while another noted employees ``scolded for not complying with record keeping policy.''
Some reported that these issues may lessen over time (perhaps from staff turnover or gradual acceptance), but may not necessarily disappear, with a participant noting ``People trying to evade or push back against record keeping was a huge one in the beginning, and some people still try to dodge it sometimes.'' In some cases, resistance may be for personal benefit, with one participant reporting ``there have been instances of people performing mass changes to information so that it fits their KPI requirements before reverting them back after audits are done.''} Such forms of `everyday resistance' ~\citep{scott1985}  (typically concealing the resistor or their activities) may be informal and small-scale , but could become a wider pattern of resistance that challenges the overall accountability regime.

\subsubsection{(5) Elaboration.~ } \label{sec:survey-elaboration}

The final step of Agre's `capture' involves storing and analysing records of the captured activity made suitable for computerised tracking. Through the accountability capture lens, \textit{elaboration} includes auditing, analysing, and adjusting record-keeping apparatus built around algorithmic systems for various purposes.

\textbf{Record-keeping is utilized for many accountability purposes, including both internal monitoring processes and compliance with legal and regulatory requirements.} The most commonly selected responses suggest that, {at present,} organisational (i.e. \textit{internal}) performance monitoring processes were the predominant reason for record-keeping. 
74\% of respondents said that the organisations kept records to monitor or improve the system's performance, while 65\% said that records were kept to verify correct system operation or identify bugs, issues, and failures (Fig. \ref{fig:records-reasons}). In a similar vein, 53\% and 51\% of respondents said records were kept for support purposes and to verify appropriate use, respectively. The key purposes for keeping records are relevant both towards the design of the capture apparatus, as well as when considering the risks for data re-use towards purposes not initially outlined during record-keeping analysis to instrumentation. {As external accountability regimes establish more explicit record-keeping requirements, new requirements may be retrofitted or layered on top of record-keeping mechanisms established for internal regimes.}

\textbf{The \textit{elaboration} stage also reveals that imposed and instrumented record-keeping practices may require re-configuration.} 16\% of organisations identified issues in their records, including missing details, irrelevant or over-recorded information, and irreconcilable system inconsistencies. 11\% of participants had also received queries or challenges from outside organisations regarding their records, including client or regulatory body activities.  Accountability capture thus risks producing `ripple effects'~\citep{selbst2019Fairness} which can alter the organisational contexts where record-keeping practices are implemented, potentially undermining their goals. In spite of these potential effects, only 18\% of participants agreed that potential for data re-use was a concern for their record-keeping practices. Thus, our results indicate that participants, and by extension, their organisations may not be sufficiently informed or supported to critically reflect on their record-keeping practices or recognise the effects of accountability capture.

\subsection{Summary}

In all, our survey of practitioners found evidence for accountability capture's processes and effects in practice. This showed that \textit{(i)} the vast majority of respondents' organisations employed various forms of record-keeping; \textit{(ii)} both \textit{internal} and \textit{external} accountability regimes drive record-keeping; and \textit{(iii)} that record-keeping does produce changes in employees' behaviour --- both in accordance with or to evade the record-keeping apparatus. 
These findings draw attention to the under-considered issues which organisations face in keeping records about their activities in line with accountability regimes (whether imposed internally or externally), and underscore the importance of considering accountability capture effects when developing and implementing laws, regulations, standards, and practical record-keeping processes for algorithmic systems.

\section{Discussion} 

{Record-keeping is widely embedded in organisational processes, yet takes diverse forms in practice.} Internal accountability regimes widely drive record-keeping for system monitoring, verification, and support, with external regimes including legal/regulatory drivers across sectors influencing the design of associated mechanisms. These systems are also often deployed across sectors subject to their own unique regulations. Given that our survey results identified changes in human behavior enabled by accountability capture, it is important to examine record-keeping contexts in other key sociotechnical research areas. We now discuss our results in the context of (1) surveillance and (2) privacy\slash data protection to explore real-world effects of accountability capture and better place communities, including practitioners, regulators, and researchers, to understand and mitigate related concerns during record-keeping design and implementation.

\subsection{Surveillance Considerations}
\label{sec:surveillance}

Accountability capture naturally facilitates surveillance of actors and of organisations and individuals who use the actors' products and services. In study of digital technologies, however, surveillance (connected to \textit{`visibility'}~\citep{foucaultDP}) can be understood more broadly to involve the collection and processing of information to influence or control people~\citep{lyon2001}. This is typically through targeted interventions to alter their circumstances or opportunities (often through `\textit{social sorting}'~\citep{lyon2003}), or by producing behaviour changes due to one's awareness that they may be `watched' at any given moment, thereby applying rules and norms themselves without requiring further direct intervention (a process known as `\textit{discipline}'~\citep{foucaultDP}). We therefore recognise {accountability} capture as a process of surveillance of people and organisations.

Accountability regimes operate both through a process of discipline, as actors subject to an accountability regime align their behavior with imposed requirements under knowledge that their recorded activities may be analysed {(see \S\ref{sec:survey-instrumentation})}, \textit{and} through targeted interventions imposed by forums to change actor behavior after record-related analysis.

\subsubsection{Visibility \textit{Over} Organisations.}

Accountability capture can afford visibility \textit{over} organisations by enabling record-keeping that reveals information about their internal systems and processes. This affords surveillance of organisations responsible for algorithmic systems and processes (or, \textit{sous}veillance or `undersight'~\citep{mann2003} where those subject to power and control by organisations collect information to understand how they operate). Facilitating this kind of sur-\slash sousveillance to understand what actor organisations are doing (and how) is often the point of accountability-related discussions and interventions. As such, such sur-\slash sousveillance may generally a beneficial and intended outcome in accountability regimes. Yet, this also brings further surveillance-related considerations in two areas: surveillance of actor organisations themselves, and surveillance of employees of those organisations.

\textit{Corporate surveillance.}
~
We have discussed how accountability regimes generally aim to make visible the systems, processes, and activities of actors. This visibility may raise concerns about the protection of intellectual property, for instance, as organisations provide information about their processes to external bodies or end-users~\citep{singh2019sec}. {Indeed, several survey respondents indicated such concerns: ``if the data was leaked,'' said one, ``it could help our competitors and our clients competitors.'' Another respondent was worried that records about their organisation's activities could be ``leaked by accident and/or used [by] the [cloud] service providers for their purposes.''} While such concerns are understandable from a corporate point of view, {and thus may directly impact internal accountability regimes}, they may be overridden by {justifications for record-keeping arising from} overarching public policy goals, rights, or other reasons, and to facilitate litigation against them.

\textit{Employee surveillance.}
Visibility over actor organisations often entails surveilling employees --  likely an intended outcome of many accountability regimes and may be desirable in high-stakes domains (e.g., pilots). However, accountability capture may enable employee surveillance which is not desirable, necessary, or proportionate (\S\ref{sec:survey-instrumentation}). 47\% of respondents indicated that records were used ``to monitor or improve staff performance'' (Fig. \ref{fig:records-reasons}), e.g.  ``keeping an eye on employees" to ``improve [their] performance,'' and ``to track in case employees go rogue.'' Some reported record-keeping was  to bring performance improvements, but others commented on employees being ``less active with their work", or becoming ``depressed or less focused," due to  the ``Orwellian feeling" brought about by record-keeping.

Employee surveillance is particularly concerning given the increasing metricisation of work~\citep{edwards2018}, whereby many aspects of working life are quantified, reconfigured, and subjected to performance monitoring and targets (often through a process of capture) such as  Amazon's retail workforce surveillance ~\citep{williams2021bi}. Those in low paid or precarious work are particularly at risk, disproportionately including women and people of colour~\citep{oneil2016}. As a result, organisations, comprised of many actors including legal departments, quality assurance, compliance teams, product managers, and C-suite executives (\S\ref{sec:survey-imposition}), may {analyze disparate impact on and implement stakeholder engagement processes} with affected communities %(e.g., employees or users monitored by algorithms). 
They might also institute policies to restrict records about employees being used for worker surveillance, or have processes to regularly check that the `grammars of action' underpinning record-keeping practices properly describe the activity and reflect the accountability aims.

\subsubsection{Visibility \textit{Through} Organisations.}

Accountability capture may also afford visibility \textit{through} organisations, or insight into external people or organisations that use or are affected by the actor's algorithmic system. Our survey established the broad potential for visibility \textit{through} organisations to facilitate sur-\slash sousveillance of others using services provided by those actors (e.g., cloud services), as organisations reported recording information about users (41\% of responses), employees of other organisations (29\%), and others affected by outcomes of the algorithmic system (9\%). 
This visibility may beneficially support various accountability aims~\citep{cobbe2019} (e.g., if the organisation forms part of another's `supply-chain'). However, visibility through organisations may also potentially facilitate corporate espionage~\citep{mattioli20} (see \S\ref{sec:survey-instrumentation}),
surveillance by state agencies, or the owner, operator, or providers using such user information for their own benefit (e.g., how Amazon used sales performance records of products sold through its retail platform to identify profitable markets and launch competing products~\citep{mattioli20}). The potential visibility \textit{through} organisations created by accountability capture relates to ongoing research and broader calls for transparency and accountability given the opaque nature of many AI and/or data-driven supply chains \cite{cobbe2021, Widder_Nafus_2023}.

\subsection{Privacy and Data Protection Considerations}
\label{sec:privacy}

Privacy relates to questions of what happens with information about people (\textit{personal data}), who has access to that information, making sure information flows are contextually appropriate, etc. 
Privacy brings practical tensions, in meeting the needs of accountability regimes, and respecting the wishes, expectations, and legal rights of those whose information is recorded or shared with others.
Understanding the process of accountability capture and its effects can help legislators consider the impact of their regulations on actor organisations and subjects affected by algorithms (i.e., visibility afforded through organisations).

\subsubsection{Privacy by Contextual Integrity.}

Nissenbaum's \textit{contextual integrity} model of privacy recognises that norms around acceptable collection, sharing, and use of information depend on 1) \textit{norms of appropriateness}, governing what information is contextually appropriate to record, and 2) \textit{norms of flow or distribution}, governing with whom it is contextually appropriate to share information~\citep{nissenbaum2004}. Our survey revealed that strategies to establish and shape \textit{norms of appropriateness} for automatic record-keeping included 1) stakeholder engagement (\S\ref{sec:survey-analysis}) and 2) communication with users, whereby an ``organization needed to clearly communicate why this data was being collected, how it would be used to provide value, and how their privacy and rights would be protected.'' 
\textit{Norms of flow} 
relate to which forums can access and view accounts by actors. For example, GDPR enables regulators to access personal data processing records  [20, Art. 30, recital 82], which may violate user norms of flow (see e.g.  \cite{Bauer_Gerdon_Keusch_Kreuter_Vannette_2022}). 
Thus, the needs of accountability regimes must be balanced with respecting the privacy (as contextual integrity) of those whose data may be recorded.

\subsubsection{Data Protection.~ }
Record-keeping and the related reconfiguration of systems may also involve processing personal data within the scope of data protection law~\citep{dphandbook}. Some survey respondents noted priorities ``to make sure nobody accesses the recorded data because it would lead to (potential) fines'' and of ``keeping the personal information of customers safe.'' The predominance of organisational accountability regimes as drivers for existing record-keeping is not surprising (\S\ref{sec:survey-elaboration}) given little direct regulation on algorithmic record-keeping. Still, 31\% of respondents said record-keeping was undertaken for legal and regulatory compliance purposes, with data protection laws cited as a primary driver. {As one respondent put it, ``a lot of GDPR and other regulations require endless retention of business activity records.'' Requirements in legal and regulatory accountability regimes may override organisations' preferences on record-keeping (see Fig. \ref{fig:records-likert}), as some respondents indicated obligations to record information they would prefer not to  and ``because of the law it's kind of concerning to keep all the data." As future regulations target algorithmic accountability, these sentiments may become common.

\textit{Organisational mechanisms.~ } Data protection principles may affect what kinds of records organisations keep, for how long, and for what purpose. Organisations may be inclined to `cast the net wide,' or capture large amounts of information to ensure sufficiently available records, and later restrict the scope of record-keeping as more is learned about the system and domain. However, this `over-collection' of data may not reflect common (cross-jurisdictional) data protection principles of \textit{purpose limitation} (keeping and using records only for specific purposes), \textit{data minimisation} (recording only the information needed for the given accountability regime), and \textit{storage limitation} (keeping records only for as long as it is needed), which will be relevant for organisations designing and implementing record-keeping systems.

\subsection{Technical Implementation Considerations}
Accountability regimes, whether internally or externally imposed, can also bring specific technical and practical considerations. Record-keeping processes can entail added functionality that introduces time- and scale-constrained overheads to `real-time' systems, as well as data storage requirements for appropriately capturing all relevant actions within a system ~\citep{camflow} (e.g., mobile apps generating gigabytes of audit data in minutes~\citep{cloete2021}). 
Record storage location and format also invoke associated risks, as choices such as offloading data to the cloud might assist scaling but raise security, privacy, legal and other concerns. The dependencies (supply-chain) supporting record-keeping also bear consideration, as emphasised by the vulnerability in the popular logging utility \texttt{\small log4j} placing many dependent systems at risk~\citep{log4j}.
Moreover, mechanisms deployed to mitigate the risks of accountability capture \textit{themselves} must also be assessed as to their potential impacts. For example, desensitising data might hinder types of oversight, such as where deleting images of individuals not recognised (for `privacy') hindered fairness assessments of a problematic police face recognition 
system~\citep{bridges}. {As such, future work may investigate the impacts of technical implementation choices made during the (re-)orienting of record-keeping mechanisms to support accountability regimes.
}

\section{Conclusion}
Calls for greater accountability encourage increased record-keeping surrounding algorithmic systems. In accountability regimes, record-keeping is often the means through which transparency is realised, supporting oversight, investigation, contestation, and redress. \textit{Yet, the impacts of these record-keeping practices remain under-examined}. Here we describe \textbf{accountability capture}, which considers the re-configuration of socio-technical processes and associated downstream effects relating to record-keeping for algorithmic accountability. From a practitioner survey, we reveal record-keeping practices and indicate evidence of accountability capture, including positive and negative behavioural changes from those subject to these processes. Our work argues an \textbf{urgent need to critically evaluate record-keeping requirements and practices, along with their implications}. 
Accountability capture shows how implementing record-keeping for transparency can reshape socio-technical systems and accountability regimes, revealing broader and potentially unintended consequences.

\section*{Acknowledgments}

We acknowledge the financial support of UK Research \& Innovation (grants EP/P024394/1, EP/R033501/1), The Alan Turing Institute, and Microsoft, through the Microsoft Cloud Computing Research Centre. SC is a PhD student in the NIH Oxford-Cambridge Scholars Program. 

\bibliography{referenceslong}

\setcounter{figure}{0} 
\renewcommand{\figurename}{Supplementary Figure}

\begin{figure*}[h]
  \begin{subfigure}[t]{0.48\textwidth}
    \includegraphics[width=\textwidth]{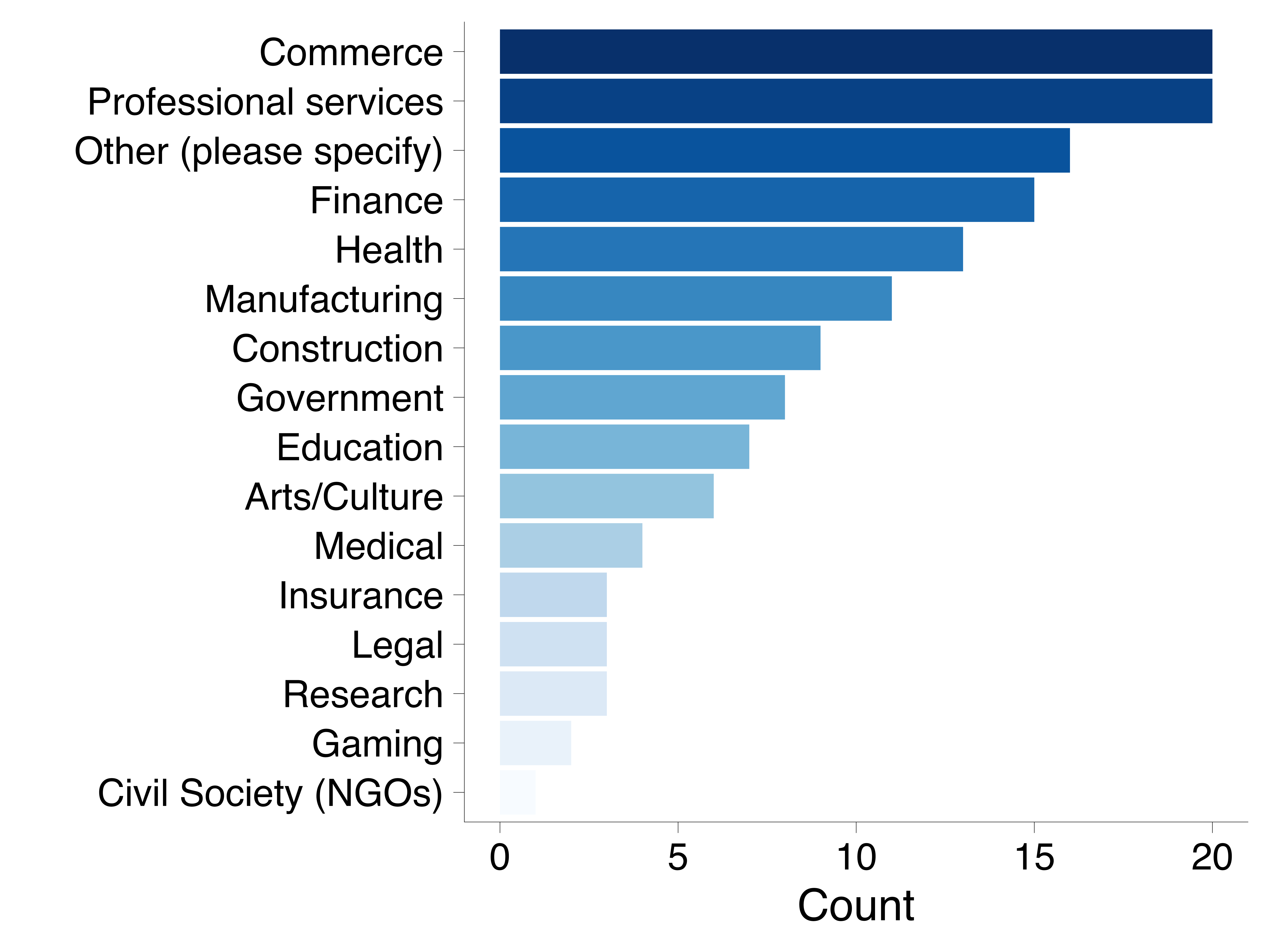}
    \caption{What sectors does your organization primarily focus on? Check all that apply.}
    \label{fig:participant-sectors}
  \end{subfigure}
  \hfill
  \begin{subfigure}[t]{0.49\textwidth}
    \includegraphics[width=\textwidth]{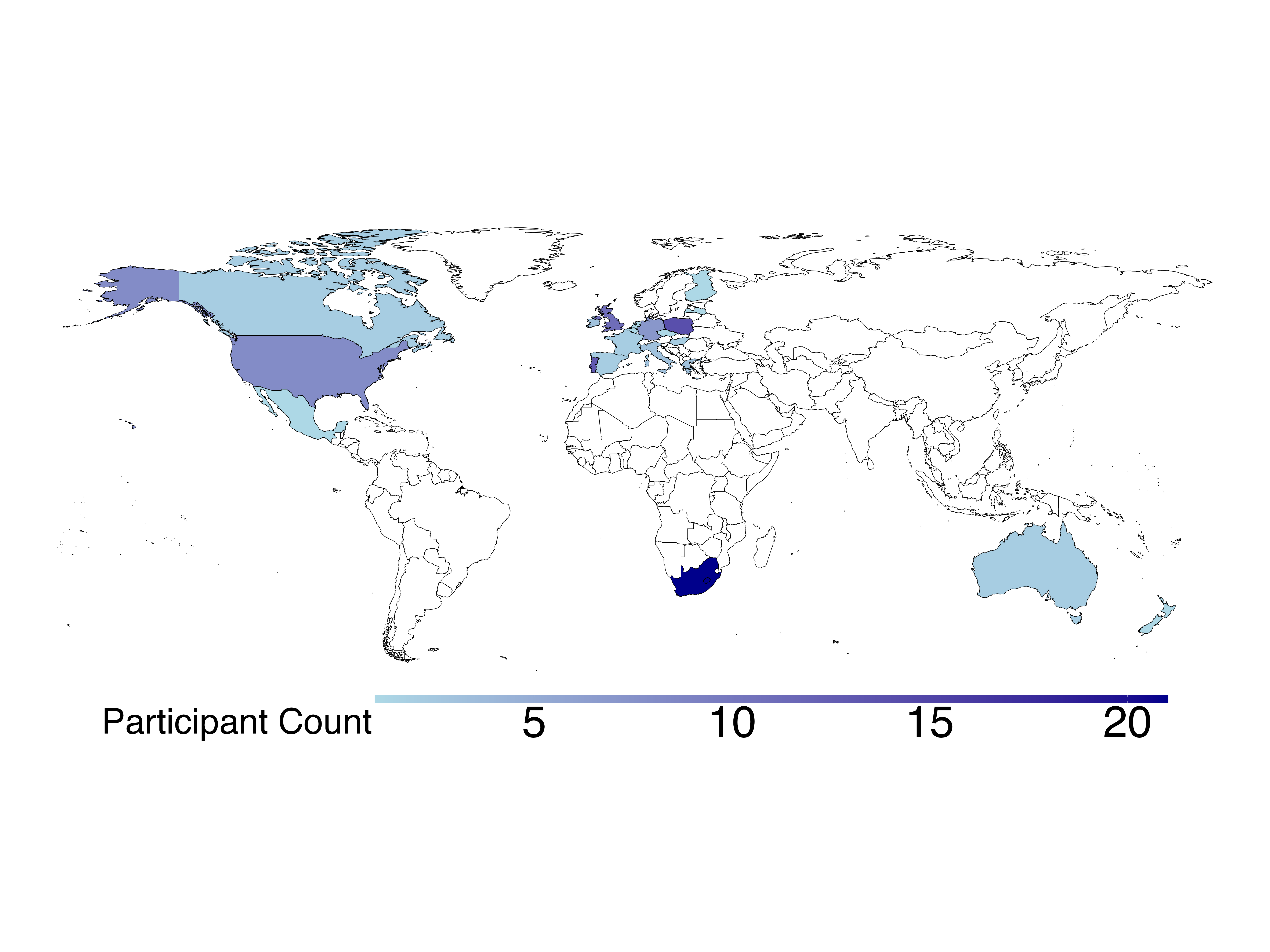}
    \caption{What country are you from?}
    \label{fig:participant-countries}
  \end{subfigure}
  \begin{subfigure}[t]{0.49\textwidth}
    \includegraphics[width=\textwidth]{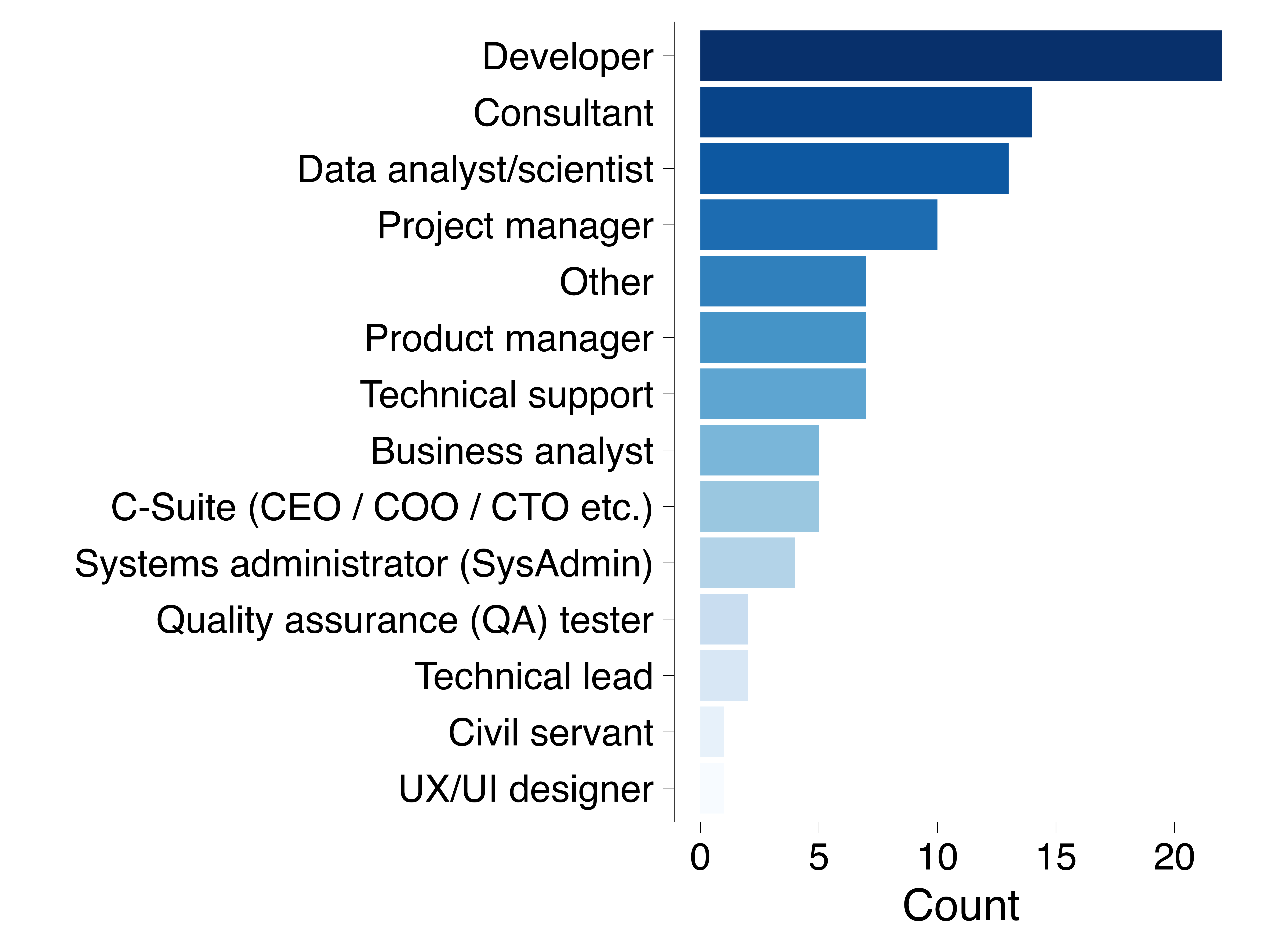}
    \caption{How would you describe your role?}
    \label{fig:participant-roles}
  \end{subfigure}
  \hfill
  \begin{subfigure}[t]{0.49\textwidth}
    \includegraphics[width=\textwidth]{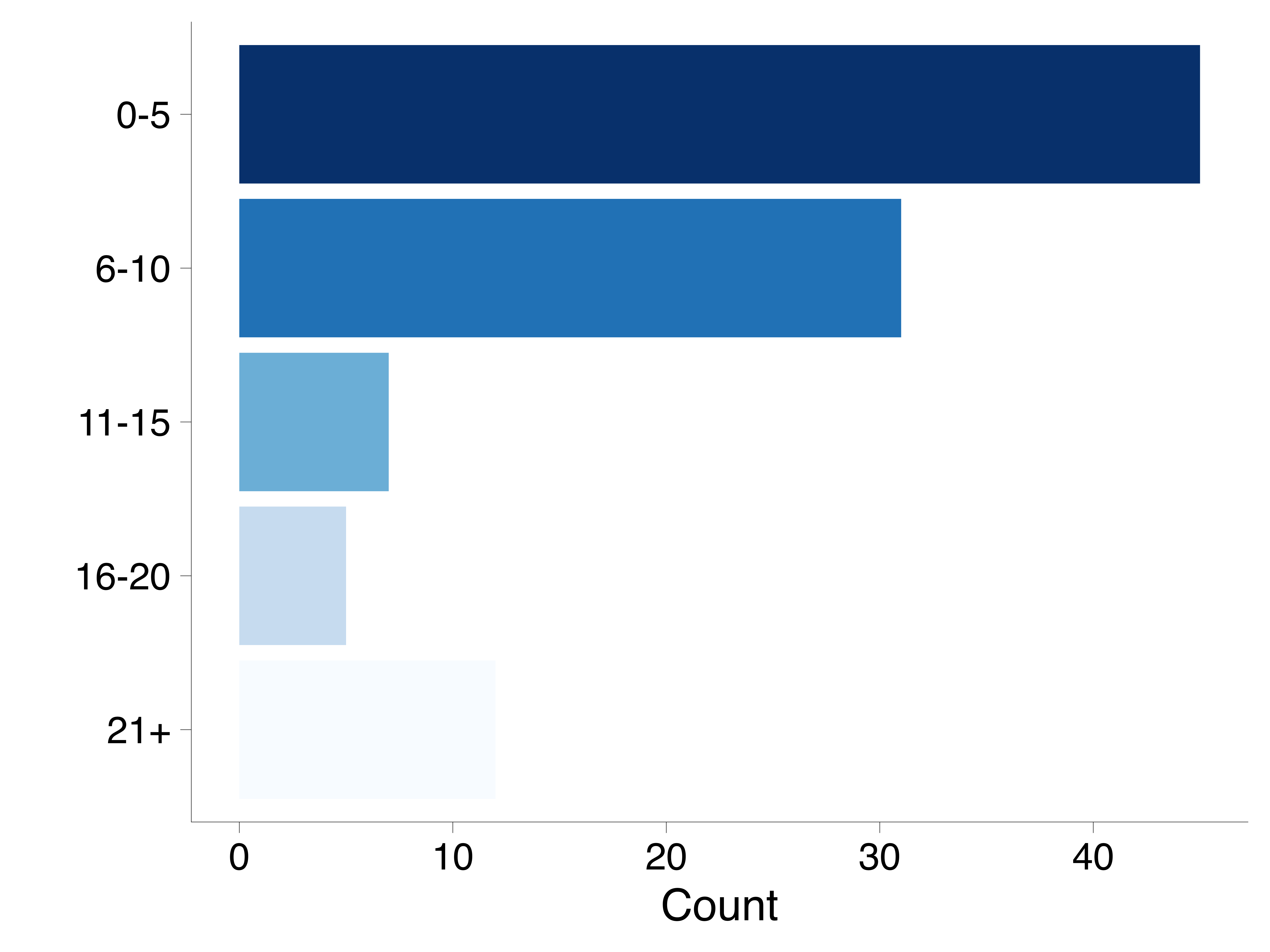}
    \caption{How many years of progression experience do you have?}
    \label{fig:particpant-years-experience}
  \end{subfigure}
  \caption{Reported participant demographics, with question text described in captions.}
  \label{fig:participant-demographics}
  % \supplementarysection

\end{figure*}

\end{document}